# 4D Printing of Programmable Digital Metamaterials

Ido Levin[1,2], Ela Sachyani[3], Rama Lieberman[3], Noa Battat[3], Eran Sharon[1], Shlomo Magdassi[3]

## Abstract

Advances in 3D printing technology now enable the precise positioning of microscopic material voxels to form complex structures. Combined with emerging multi-material capabilities and printable responsive materials, this opens new possibilities for digital composite materials and 3D printing of shape-transforming structures, or 4D printing. Building upon these advancements, we devise a novel methodology for crafting digitized 4D-printed shape-transforming sheets. We 3D print responsive sheets composed of two layers, each consisting of active and passive voxels meticulously positioned to form thin structures that can be actuated on demand. Our approach solves a long-standing problem in the field, i.e., the independent and simultaneous programming of lateral geometry and reference curvature. This unprecedented control over the resulting shape unlocks new opportunities in synthetic shape-morphing materials, with potential applications in programmable mechanical properties and multi-material systems.

## Introduction

The striking ability of many thin biological structures to induce morphological changes relies on differential swelling – a non-uniform swelling of the structure [1–3]. The evolving shape is fully controlled by the local information encoded within the swelling magnitude and orientation at different locations. For thin sheets, differential swelling consists of two complementary orthogonal modes: the *bilayer effect* and *Gaussian morphing*. Gaussian morphing involves inducing lateral deformation gradient to program Gaussian curvature, allowing, theoretically, to realize any 2D surface. However, there is no control over the sign of surface bending (upward or downward). The bilayer effect involves inducing deformation gradients across the thickness of the sheet, programming spontaneous curvature. This provides control over the sign of the curvature but is limited to uniaxial response, i.e., surfaces with zero Gaussian curvature.. Therefore, to achieve complete control over the actuated shape, one must be able to impose *both effects simultaneously*. In other words, accurately inducing deformation gradients in 3D - both laterally and across the thickness.

The potential of differential swelling for applications has inspired many multidisciplinary endeavors to realize it using synthetic materials. Since small variations of swelling across the thin dimension of the sheet can induce large curvature magnitudes, most of the work was focused on programming spontaneous curvature[4–7] using the bilayer effect. Nevertheless, recently, researchers developed several techniques to accurately induce Gaussian morphing[8–11]. The actuated structures were shown to have rich configurational phase space and the capability to generate complex patterns by breaking the symmetry of the programmed swelling fields. However, up to date, the ability to determine a desired 3D shape is still only partial as simultaneous programming of both Gaussian and spontaneous curvatures has not been achieved yet.

1 Racah Institute of Physics, The Hebrew University, Jerusalem 91904, Israel
2 Department of Chemistry, University of Washington, Seattle, WA 98195
3 Institute of Chemistry, The Hebrew University of Jerusalem, Jerusalem 9190401, Israel

The principal strategy to induce differential swelling is by using responsive materials. These include hydrogels [8,11–13], nematic elastomers [9,10], and dielectric elastomers [14]. Choosing the right material involves considering the material properties and the method to program differential swelling. Hydrogels are advantageous as they can be responsive to a variety of external triggers such as temperature, pH, light, and humidity[15,16], and can be repeatedly actuated. The response can be tailored to yield a relative volumetric shrinkage as high as 100-fold [17] , and objects made of hydrogels are suitable for integration in soft robots, implants, and drug delivery systems[18–20]. Recently, there has been a growing interest in hydrogel-made objects, including fabrication by 3D printing [21–26]. Among the available 3D printing technologies, direct ink writing (DIW) and stereolithography (SLA), including digital light processing (DLP), are the most used for printing hydrogels.

Hydrogels also provide several techniques for accurate control of the swelling field. Lithography-like approaches allow very high spatial resolution[11,27]. Alternatively, shear alignment can be utilized to control the orientation of rigid fibers inside printed gel sheets, resulting in anisotropic swelling [28]. Some of these systems also enable limited gradients in the thin dimension by asymmetric polymerization[29] or by combining two different layers[30]. However, optimal three-dimensional control over the deformation field, which would enable free programming of shapes, has not yet been achieved.

In this work, we present a novel 3D printing strategy for fabricating shape-morphing hydrogel sheets with fully programmable 3D swelling gradients. As such, it allows the actuation of both morphing modes simultaneously, resulting in smart materials with a broad range of programmable 3D shapes. Our approach to printing swelling gradients is by digitizing the smooth swelling field and printing voxels with different swelling ratios. This is analogous to how a regular printer prints grayscale gradients by printing black and white pixels. As such, it provides a general method to fabricate multi-material self-morphing structures via pixelization, a method that is applicable to many printable materials. Moreover, it can be easily extrapolated for materials with anisotropic swelling response, such as nematic elastomers. Specifically, we fabricate bi-material composite structures with an SLA 3D printer. Both materials are polyacrylic acid polymers crosslinked with polyethylene glycol diacrylate (PEGDA) but have different crosslinking densities. This enables printing two voxel types that swell by different ratios upon hydration. The relative spatial density of the printed voxels of each type determines the magnitude of the local swelling, just as the density of black pixels determines the grayscale value. In a coarse-grained view, it enables programming continuous 3D swelling fields. We repeat this process for two layers, resulting in high-resolution digital 4D structures with a swelling factor that can be programmed and controlled in 3D, thus locally determining both the spontaneous and Gaussian curvatures. Therefore, our method can induce both shape-morphing modes simultaneously.

The shaping of thin sheets via differential swelling can be formulated in the theory of non-Euclidean elasticity [31]. In this formalism, the differential swelling is represented by geometrical quantities: lateral swelling gradients are encoded in the *reference metric,* $\bar{\boldsymbol{a}}$, and gradients across the thickness are encoded in the *reference curvature,* $\bar{\boldsymbol{b}}$. The programmed Gaussian curvature can be directly calculated from $\bar{\boldsymbol{a}}$ (see supplementary information). In many cases, $\bar{\boldsymbol{a}}$ and $\bar{\boldsymbol{b}}$ are incompatible; that is, no surface can simultaneously satisfy both $\boldsymbol{a} = \bar{\boldsymbol{a}}$, and $\boldsymbol{b} = \bar{\boldsymbol{b}}$, where, $\boldsymbol{a}$, and $\boldsymbol{b}$, are the *actual* metric and curvature tensors of the actuated shape, respectively. In such cases, the shape is determined by a competition between bending and stretching. This competition is the source for many of the rich phenomena observed in such systems: shape transitions [4], breaking of symmetry [32,33], fractal structures [34,35], and exotic mechanical properties[36–38]. To fully control these phenomena, one must fully control the incompatibility, again calling for complete control over the 3D swelling field.

Very thin sheets are nearly unstretchable. Hence, the configuration will obey the encoded lateral swelling gradients. In other words, very thin sheets obey the programmed Gaussian curvature. Therefore, an apparent strategy for designing a sheet that will morph into a desired shape is to program $\bar{\boldsymbol{a}}$ that encodes the Gaussian curvature of that surface, a problem that can be solved analytically[11] or numerically[39]. However, finding the metric is not enough as there usually are many different surfaces with the same Gaussian curvature but a different shape, and therefore, guiding the shape selection using $\bar{\boldsymbol{b}}$, *in addition* to $\bar{\boldsymbol{a}}$, is crucial[13]. For instance, programming a constant positive Gaussian curvature (corresponding to a spherical cap) will result in a bistable structure that can buckle upward or downward. The spontaneous curvature must also be determined to assure buckling upward into a dome or downward into a cap. Indeed, the impressive control over $\bar{\boldsymbol{a}}$ presented in [13] and [40] could guarantee the desired 3D shapes only in a limited range of cases. Moreover, when the sheet is not infinitely thin, finite thickness effects appear, resulting in a deviation from the programmed Gaussian curvature. Finally, beyond full control over shaping, the ability to simultaneously control both $\bar{\boldsymbol{a}}$ and $\bar{\boldsymbol{b}}$, could open the way for programming exotic mechanical properties into elastic sheets [41].

## Results

The new approach is based on three stages: 1. *Design*, in which the smooth swelling field is calculated. 2. *Pixelization*, in which the smooth field is digitized into the desired resolution. 3. *Fabrication*, in which the desired flat structure is 3D printed (Fig. 1). Then, the programmed geometry is *actuated* by hydrating the gel in water.

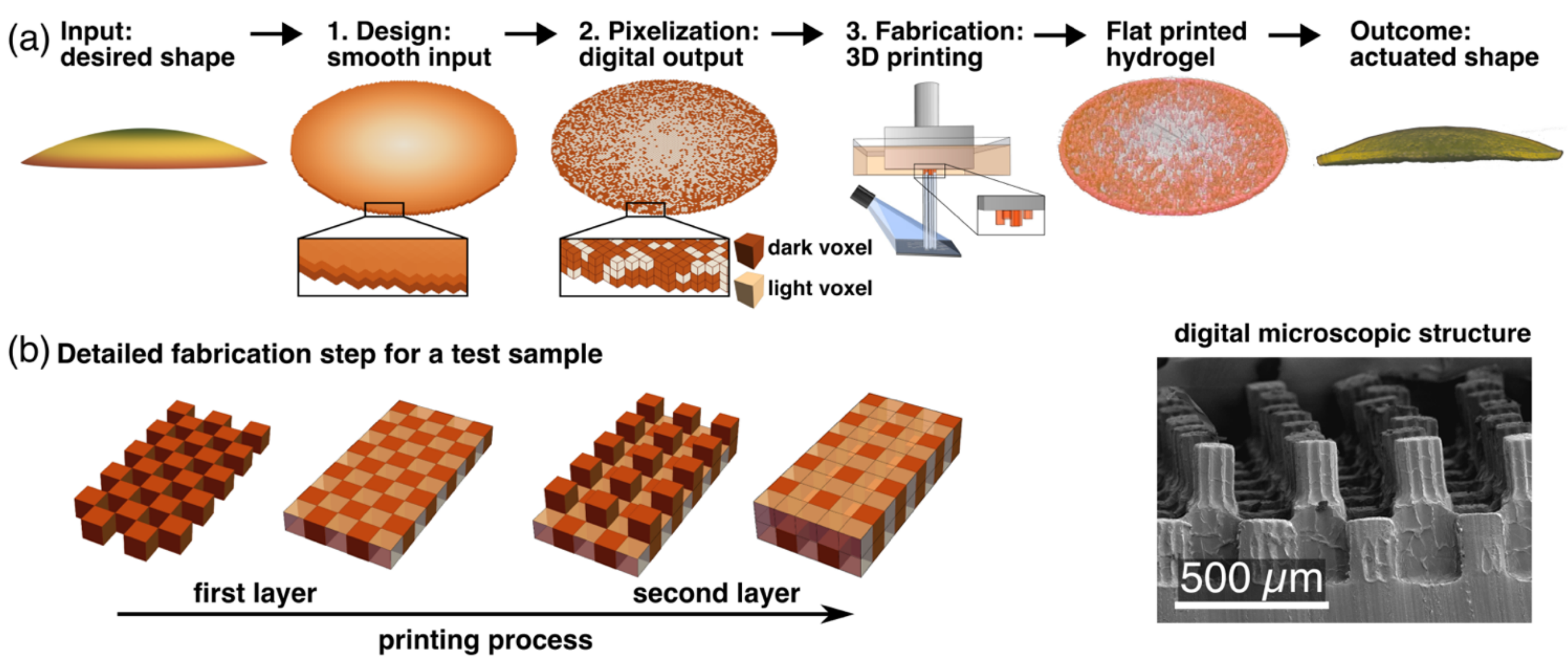

*Figure 1.* ***Design, Digitization, and Fabrication**. (a) Outline of our technique: The geometry of the desired shape is analyzed, resulting in its metric and curvature fields. The analysis results in a smooth conformal swelling density field for each layer(1). The two smooth density fields are digitized(2). The digitized geometry is realized by multi-material 3D printing(3). The resulting flat sample is actuated in water, forming the desired 3D shape. (b) Detailed Illustration of the fabrication step. A test voxelated bilayer structure consisting of two types of dark voxel stacks: 200μm (a single voxel) and 400μm (two stacked voxels). Steps of the multi-material printing (left). SEM image of the printed structure (here, only the dark voxels were printed for visualization; right).*

While this approach has the potential to print fully 3D structures, here we focus on printing thin structures. Moreover, to actuate both morphing modes, two layers are necessary and sufficient. Hence, we print bilayer structures, allowing simultaneous control over both $\bar{\boldsymbol{a}}$ and $\bar{\boldsymbol{b}}$. The relative densities of the two voxel types dictate the local effective swelling ratio. In other words, the discrete nature of the voxelated structure, while smoothed by the elastic response of the material, yields an effective smooth reference geometry. Therefore, an important quantity is the ratio between the thickness and the spatial resolution of the voxels. This ratio must be large enough, i.e., the voxel size must be small compared to the thickness [11] (similar to how black and white pixels are converted into a grayscale image when the resolution is sufficiently high). When this ratio isn't large enough, the sheet can undergo local buckling at the voxel scale. In those cases, the actuated shape will deviate from the programmed one. Benefiting from the high resolution of typical commercial DLP printers, we can make sure that the digitization is made in a length scale that is smoothed by elasticity, reproducing the original continuous field. Indeed, the thickness of our samples is twice the voxel size and, therefore, well within the homogeneous range found in [11] .

Each printed layer is formed as follows: voxels of the first material are printed, the bath is replaced, and then a layer of the second material is printed (Fig. 1b). For visualization purposes, we color orange the voxels made of the gel with the higher crosslinking ratio. The colored and uncolored voxels are referred to as "*dark*" and "*light*" voxels, respectively. This method is advantageous to molding the second material as it provides better control and flexibility and supports the next layer. The method is suitable for multi-material printers and can be generalized to an arbitrary number of consecutive materials. In the present work, this process is repeated twice, forming a bi-layer composite structure with the desired properties.

For each layer, the local relative spatial density of the two voxel types provides an effective swelling field. Denoting these swelling fields at the top and bottom layers, $\omega_t$, and $\omega_b$, respectively, the resulting non-Euclidean reference tensors read (see supplementary information for details):

$$(1)\ \bar{\boldsymbol{a}} = \frac{1}{4}(\omega_t + \omega_b)^2 \begin{pmatrix} 1 & 0 \\ 0 & 1 \end{pmatrix},\ \bar{\boldsymbol{b}} = \frac{3}{t}\left(\frac{\omega_t - \omega_b}{\omega_t + \omega_b}\right)\begin{pmatrix} 1 & 0 \\ 0 & 1 \end{pmatrix}$$

Note that while an isotropic swelling provides full control over, $\bar{\boldsymbol{a}}$ , (i.e., any reference metric can be realized), it can only prescribe isotopic curvatures. Nevertheless, it controls the *sign* of curvature, which is impossible by prescribing only $\bar{\boldsymbol{a}}$.

Since the printed sheets are thin, they are almost unstrachable. This implies that to obtain the desired configuration with some metric $\boldsymbol{a}$, $\omega_t$ and $\omega_b$ must be properly selected to have $\bar{\boldsymbol{a}} \approx \boldsymbol{a}$. By that, we also overcome the limitation of our ability to control $\bar{\boldsymbol{b}}$, as $\bar{\boldsymbol{a}}$ induces the Gaussian curvature and $\bar{\boldsymbol{b}}$ controls the mean curvature (by biasing the curvature sign). Therefore, we start by computing the metric first, and then adding the curvature to each layer. In this case, it is more natural to express Eq. 1 using the *metric generator* (average swelling), $\omega_0 \equiv \frac{1}{2}(\omega_t + \omega_b)$, and the *curvature generator* (swelling variation) $\kappa_0 \equiv \frac{3}{2t\omega_0}(\omega_t - \omega_b)$, which reads:

$$(2)\ \bar{\boldsymbol{a}} = \omega_0^2 \begin{pmatrix} 1 & 0 \\ 0 & 1 \end{pmatrix}, \bar{\boldsymbol{b}} = \kappa_0 \begin{pmatrix} 1 & 0 \\ 0 & 1 \end{pmatrix}$$

, where, $t$, is the thickness of the printed sheet.

For simple cases, the metric generator can be analytically calculated directly from the desired shape (see supplementary material). For more complex shapes, one can use numerical methods such as the Boundary First Flattening (BFF) algorithm that finds conformal parameterization of desired surfaces [39]. Both methods involve finding a mapping of the curved surface into a plane that does not distort angles but is allowed to swell/compress different sections (Fig. 1 a). We are still left with the freedom of selecting $\kappa_0$, to prescribe the reference curvature.

## Calibration

To calibrate our system, we must first find how the swelling depends on the density of the dark voxels, $\rho$. To achieve this, we print square calibration structures composed of a single layer with a uniform dark-voxel density (0% to 40%). The printed structures are immersed in water for swelling, up to their final size. We verify that the swollen structures remain square and flat, indicating that the printing resolution is sufficient for the voxelated structure to yield an effective uniform swelling ratio, $\omega_0$. When completely swollen, we measure the relative change in their length, $\omega_0 = \frac{\ell}{\ell_0}$, where $\ell$ and $\ell_0$ are the swollen and original lengths, respectively. As expected, the lengthening is reduced as the density of the dark voxels is increased. The resulting constitutive law is almost linear in the pillar density and presents a 25% relative longitudinal swelling between the extremal densities (Fig. 2a). This measurement enabled calibrating the emerging reference metric, hence, to inverse-design density profiles for desired shapes of plates in the thin limit.

Next, we construct sheets with uniform $\omega_0$ (hence, we program vanishing Gaussian curvature), while controlling the reference curvature by prescribing gradients in the thin dimension (controlling $\kappa_0$). We print structures with a uniform voxel density on either the top or the bottom layer, with the other layer empty of dark voxels. Upon actuation, the swelling gradient results in a uniform curvature that can be predicted by substituting the constitutive law into Eq. 2. The predicted magnitudes of the emerging curvatures match all the measurements very well, with no fitting parameters (Fig. 2b), and we demonstrate the ability to control the curvature in both directions (curving upward and downward). The uniformity of the curvature is another indication that the discrete programmed swelling field is smoothened correctly.

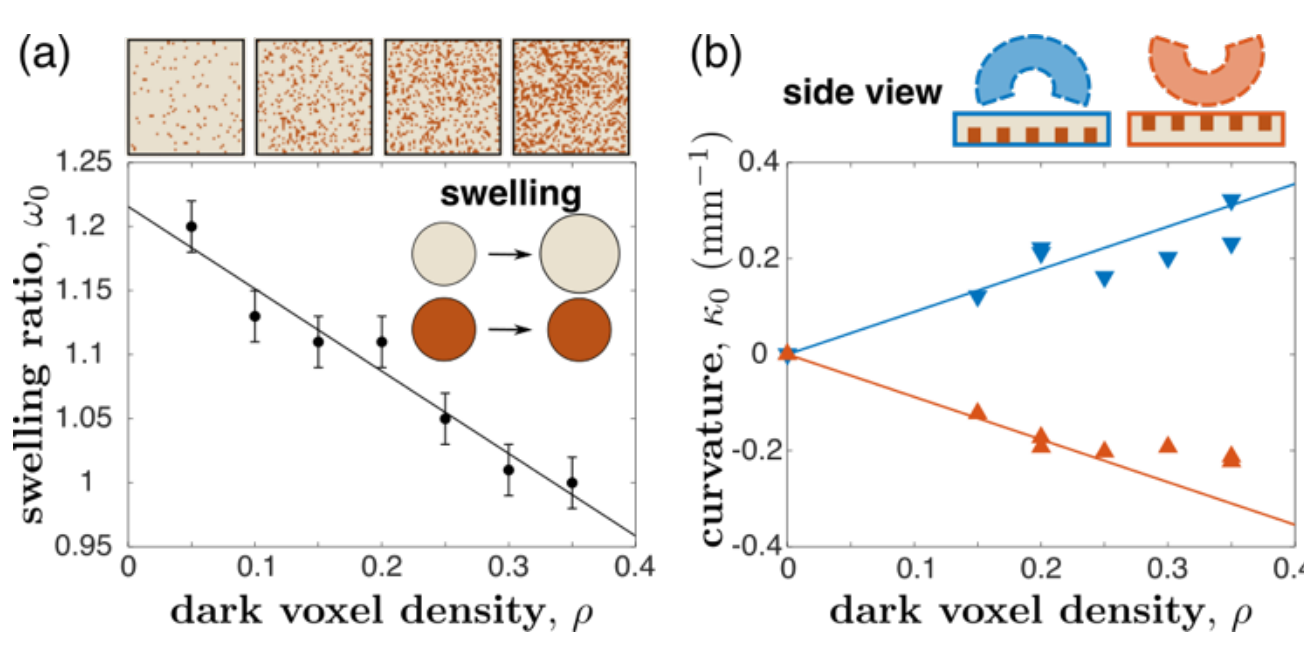

*Figure 2.. **Calibration measurement of swelling ratio and curvature**. (a) The longitudinal swelling ratio (the ratio between the actuated and original lengths), $\omega_0$, of calibration samples with a uniform voxel density (dark circles). The fitted linear trend (black line) suggests a monotonic dependence of the swelling factor on the pillar concentration. Relative swelling of almost 25% was achieved, allowing significant freedom in prescribing shapes. The variation in the voxel density is illustrated in the squares above. (b) The prescribed curvature, $\kappa_0$, for specimens with one layer with a uniform voxel density and the other with no dark voxels at all (red and blue points,*

*corresponding to dark voxels only on top or bottom, respectively). The solid lines result directly from plugging the swelling calibration fit in (a) into Eq.2 (i.e., there are no fitting parameters). These results demonstrate the capability to prescribe curvature in either direction and allow us to program the reference fields quantitatively.*

## Introducing swelling gradients

Once the constitutive law is known, we program more complicated shapes by introducing spatial gradients to the voxel densities. We start with inducing only Gaussian morphing ($\kappa_0 = 0$). Linear radial swelling profiles, $\omega_0(r,\theta) = \alpha\, r + \beta$, are sufficient to induce 3D shapes. The sign of the Gaussian curvature depends on the profile's slope: increasing/decreasing trends, prescribe Gaussian curvature with a negative/positive sign (i.e., dome-like/saddle-like), respectively. By this approach, we demonstrate the formation of 3D shapes that are governed by radial changes, such as dome-like and combined, dome-saddle-like structures(Fig. 3). The fact that the swollen structures are smooth, showing no evidence for the discretized swelling fields, is another validation that the resolution is high enough for it to be smoothened by elasticity. Our control over the shape is not just qualitative: we calculate the swelling needed for spherical caps with radii in the range 0.5-2 cm (see Supplementary Material for the calculation). Gel disks of different radii, programmed with discretized swelling fields are printed. When actuated, the disks morph into caps with the prescribed spherical angle and radius (Fig. 3 (c)).

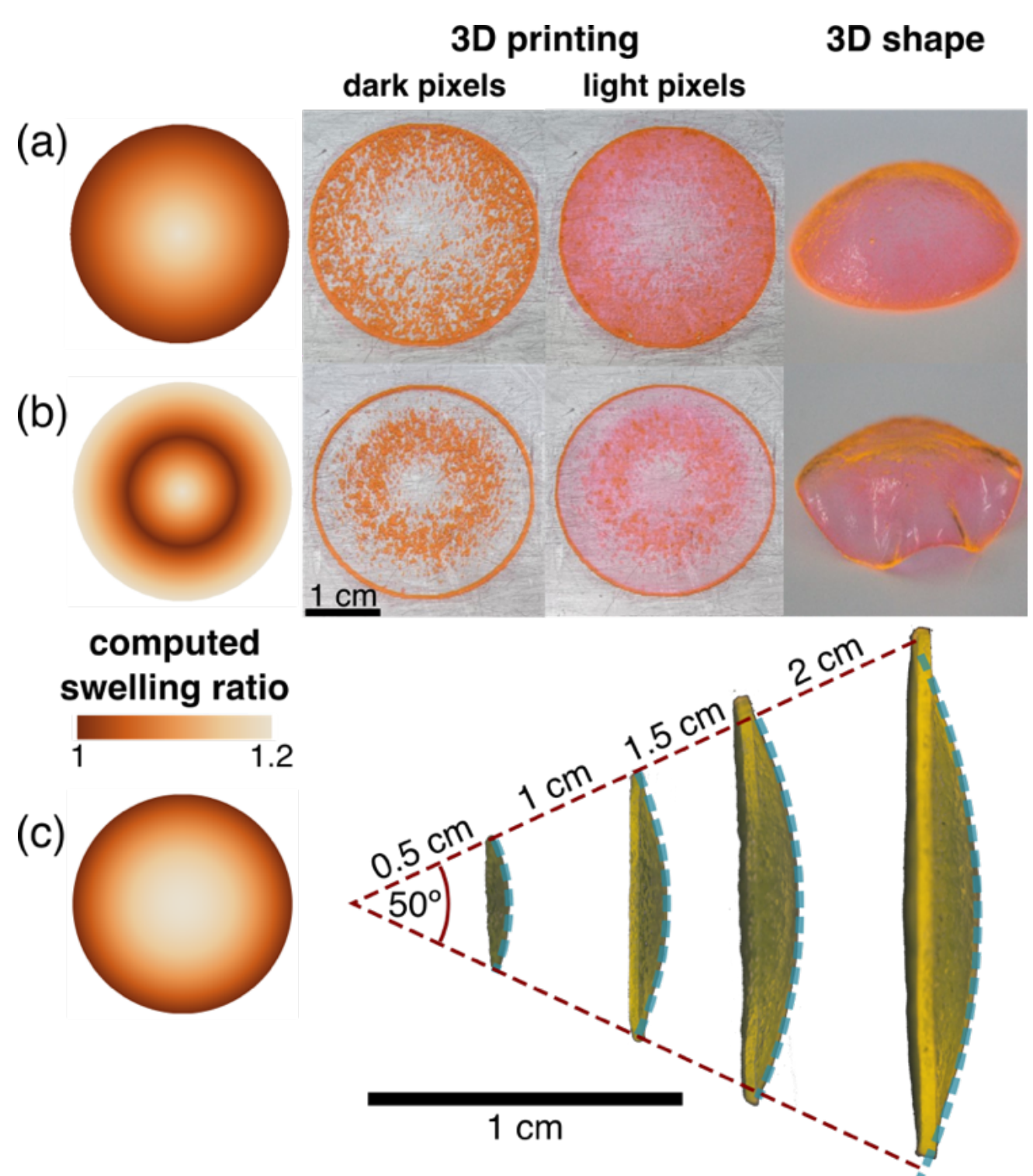

*Figure 3. **Control of Gaussian morphing by inducing lateral swelling gradients** (a-b) Simple samples with axisymmetric swelling gradients, either increasing monotonically (a, left panel), or increasing and then decreasing (b, left panel). This results in a spherical structure, or a spherical center with wavy rims (right panels of a, b, respectively). (c) Printed gels with a swelling field corresponding to the metric of a sphere. We printed four samples of different sizes with reference*

*metrics of spheres with varying radii in the range 0.5-2 cm and a fixed spherical angle of 50°(dashed teal arcs). The swollen gels adopt these configurations, as illustrated by the tight fit of the arcs.*

Next, we test the ability to actuate the bilayer effect. To this end, we 3D print disks with uniform average voxel densities (hence, a uniform $\omega_0$), and with programmed reference curvature that varies spatially. Specifically, the swelling variation varies sinusoidally between the two layers as a function of the azimuthal angle, as shown in Fig. 4. In this case, both the reference curvature sign and magnitude oscillate, bending the disk up and down. We design and print such structures with 3,4,5, and 6 oscillation nodes. The emerging shapes follow the prescribed number of nodes, indicating that we can spatially control the direction of curvature. Note that this shape has very little Gaussian curvature (the curvature is mostly uniaxial, oriented in the azimuthal direction). As such, the direction of the curvature is induced while the stretching associated with these structures remains small.

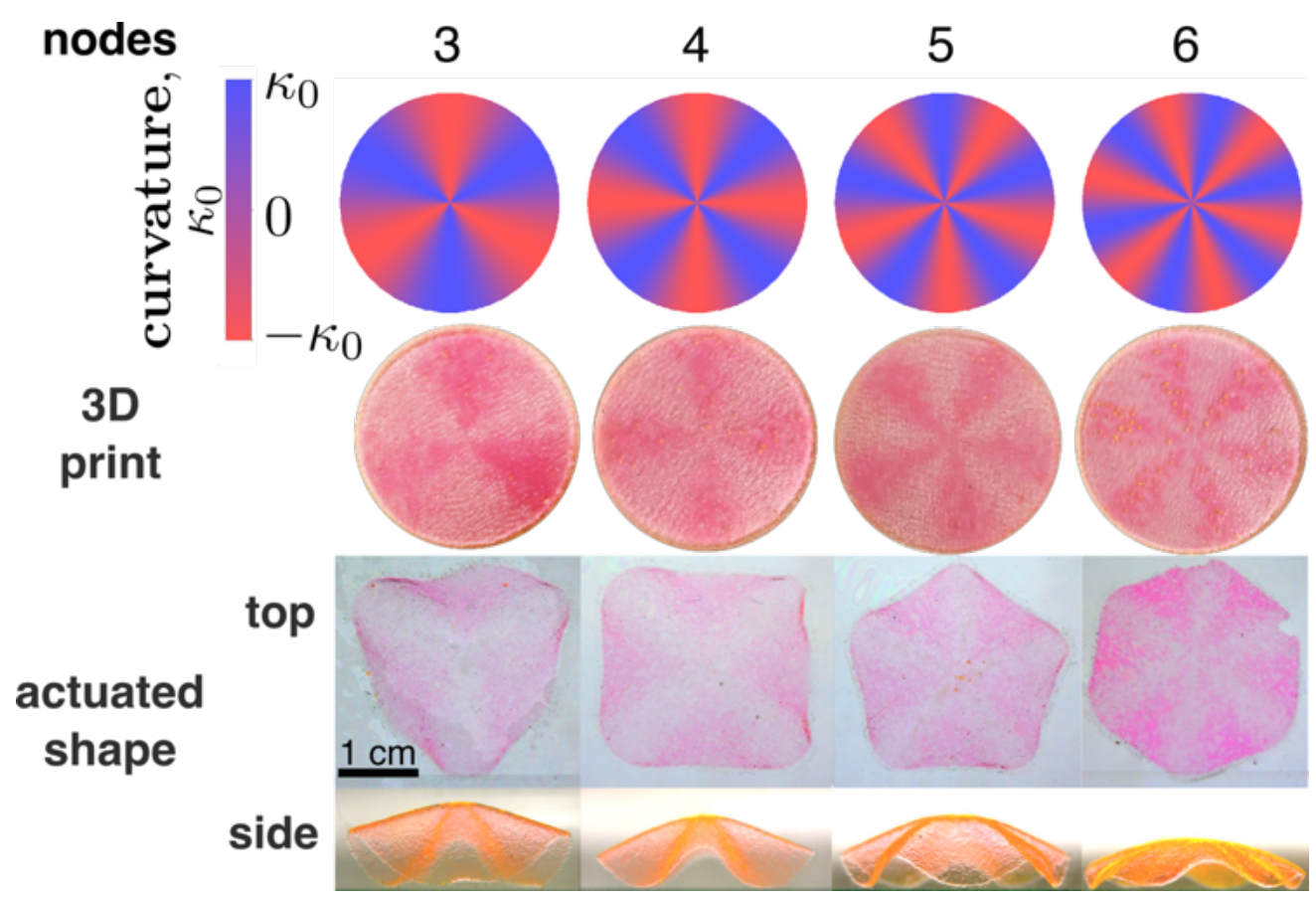

*Figure 4.* ***Control of the bilayer effect by introducing swelling gradients across the thickness****: Discs with an azimuthally oscillating reference curvature and a flat reference metric. We print structures with a varying number of nodes (3,4,5, and 6), demonstrating our ability to induce shape by controlling the spatial distribution of the reference curvature. All samples are circular. The apparent polygonal contour (third row) results solely from 3D projection effects.*

The control of either the Gaussian curvature or the spontaneous curvature, as shown in Fig. 3, 4 was demonstrated in earlier works[8,11,13]. However, the simultaneous control over these two entities, which is crucial for complete control over the shape, has not been achieved yet. We now use our ability to separately control both morphing modes to achieve this goal, i.e., to induce both Gaussian morphing and the bilayer effect, *simultaneously*. To this end, we design a model that necessitates both Gaussian curvature and spatial control over the curvature sign: two spherical caps with a small overlap (Fig. 5) curved in different directions. Each cap has the same swelling factor of a spherical cap used in Fig. 3 (c) (i.e., the same $\omega_0$ field), but a different sign of the reference curvature (i.e., the opposite $\kappa_0$ field). The magnitude of the reference curvature is designed to match the radius of curvature of the sphere, and its sign changes from positive in the left half to negative in the right half ($\kappa_0$ vanishes in the overlap region). These two conditions determine $\omega_0$ and $\kappa_0$, and hence, the entire design. *All* printed structures robustly morphed into the designed double-dome shape, with a positive and negative mean curvature in the domes. The radius of curvature of each half matches quantitively the programmed one (Fig. 5 (d)). These results display the advantage of our method, as it is absolutely impossible to get this configuration by controlling only one of the two reference fields.

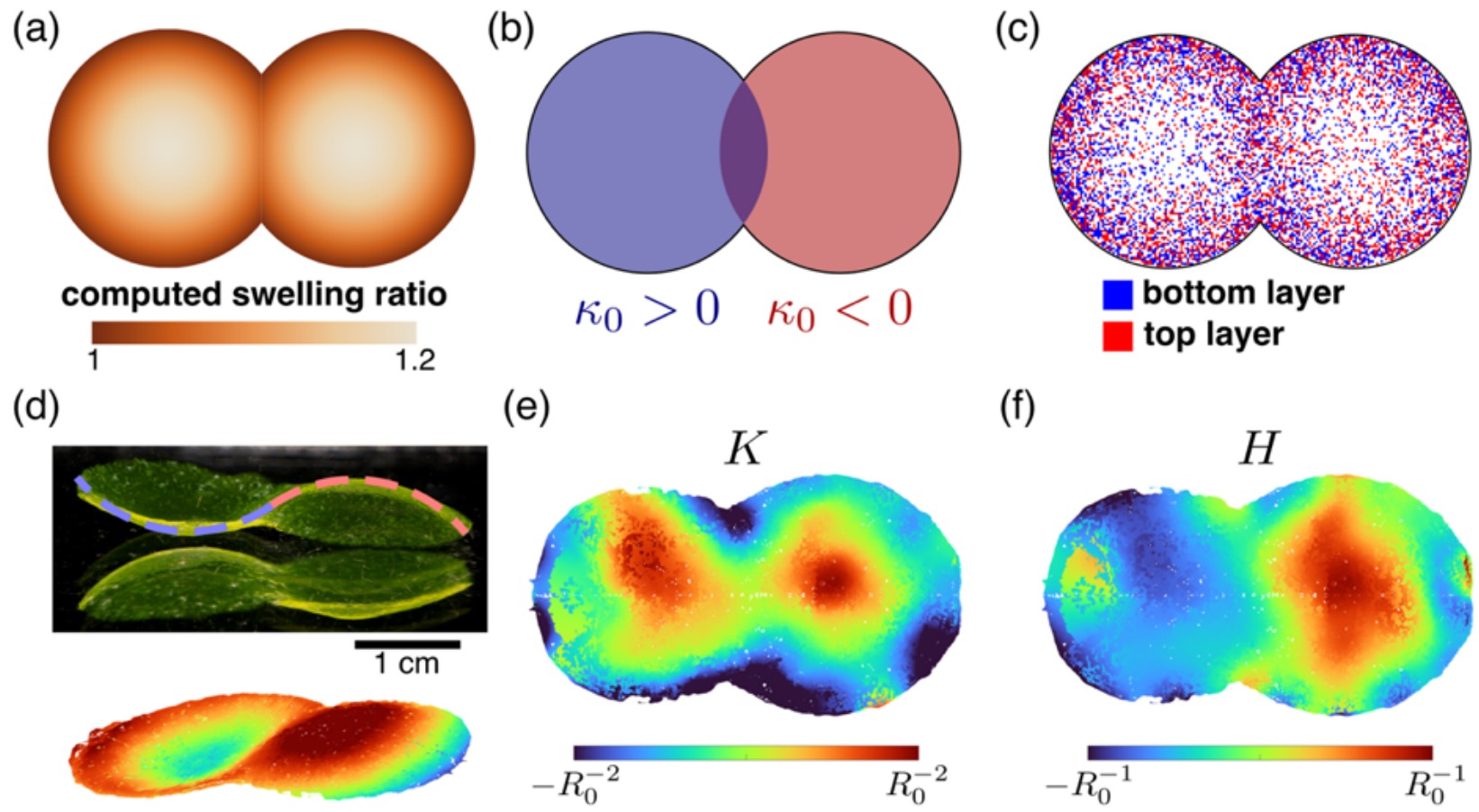

*Figure 5. **A simultaneous control over Gaussian morphing and the bilayer effect**. We design and print a structure composed of two identical spherical caps, each with a different reference curvature direction (final radius: 15 mm, printed radius: 11 mm, overlap 4.4 mm) (a) The reference metric is the same for both halves and is identical to the one on Fig. 3. (c) The reference curvature is uniform in each half and is identical in magnitude to the one encoded in the reference metric but have opposite signs. (c) The resulting voxel map, where color indicates the layer of each dark voxel (blue – top layer, red – bottom layer). (d) The swollen sample (and its reflection) demonstrates our ability to control both reference fields together (top). The dashed line corresponds to the programmed radius of curvature, illustrating the quantitative agreement with the designed configuration. The sample is 3D scanned and smoothed (bottom). The calculated Gaussian and mean curvatures, (e) and (f), respectively, illustrate the double dome structure*

## Conclusions

This work suggests a novel strategy based on 3D printing to design and fabricate self-morphing sheets of smart materials with programmed responsive geometry. Until now, experimental realizations of self-morphing were limited to the actuation of *either* the bilayer effect, or Gaussian morphing, and, therefore, provided only partial control over the configuration. Here, we go beyond the state of the art by developing a method that can fully program the swelling field, actuating both modes simultaneously. In our approach, the 3D printer is not used to build a 3D structure explicitly. Instead, we use its high resolution and accuracy to encode digital information into a thin, flat sample. Upon actuation, this information drives the self-morphing of the sheet into the desired, accurately predicted 3D shape. This strategy, i.e., of using a printer as a programming machine, the digitization, and the separation to lateral and vertical gradients are applicable to other responsive materials, suggests a new approach to self-morphing. In this work, we limited the study to isotropic swelling, which provides control over the direction of curvature. We suggest that introducing x-shaped elements (rather than circular pillars), can induce effective anisotropy to the curvature. Moreover, there is a strong desire for a shape-morphing structure that can operate under orthogonal actuation modes (such as temperature and pH), which, using traditional fabrication techniques, is not straightforward[42]. However, with our system, achieving this merely entails printing voxels composed of various responsive materials with the required actuation modes, similar to how color printing uses pixels of three colors.

Combining computation of the swelling field needed for obtaining a desired 3D shape with pixelization makes our approach compatible with all common 3D printers based on stereolithography processes, allowing one to print the structure using only two materials, thus simplifying the fabrication process. The ability to control both the metric and the curvature opens new directions for 4D designs. 3D structures can be made fully compatible (i.e., have a compatible reference metric and curvature) and thus have a single, well-defined equilibrium shape, as in Fig. 5, but can also be made weakly or strongly incompatible. As the degree of incompatibility increases, the resultant structures could have many meta-stable configurations, and the structures would become more responsive mechanically. We note that beyond the ability to specify a desired 3D shape, the ability to accurately control both metric and curvature can be used to fabricate structures with exotic mechanical properties [41], significantly expanding the notion of "smart materials". Another advantage of our strategy is the ability to extend it to different responsive material types. Here, we use hydrogels that swell uniformly with different magnitudes; however, in many cases, introducing gradients by controlling the swelling orientation (such as in nematic elastomers, for instance) is advantageous [10,28,43] . While the design stage for these materials is different[44], the pixelization can follow the same algorithm. Therefore, we provide a way to employ 4D printing for different responsive materials

## Materials and Methods

Acrylic acid was purchased from Acros (Belgium). TPO photo-initiator diphenyl(2,4,6-trimethylbenzoyl)phosphine oxide was obtained from BASF (Germany). The surfactant SDS, Sodium Dodecyl Sulfate, and dye Rhodamine 6G were purchased from Sigma Aldrich (Merck, Germany). PEGDA was obtained as a gift from Sartomer-Arkema (France). Triple distilled water (TDW) was obtained from NANOpure®-DIamondTM (TDW; 0.0055 µS.cm-1; Barnsted system, IA, USA).

***Digitazion of the concentration fields:*** Once the desired $\bar{\boldsymbol{a}}$ and $\bar{\boldsymbol{b}}$ are calculated, we extract the metric and curvature generators, $\omega_0$, and $\Delta$, and then calculate the smooth density field of each layer. Then, each voxel is assigned a material independently according to the local density value. For instance, if the local density value is 0.2, the probability that it will be assigned a dark or a light voxel, is 20% and 80%, respectively.

***Printing formulation:*** *The printing compositions are based* on materials previously developed in our lab[22]. *Two types of* UV-curable aqueous solutions inks containing acrylic acid with polyethylene glycol diacrylate (PEGDA) as a crosslinker were prepared: one with low wt% of PEGDA (2wt% PEGDA, 38wt% acrylic acid, TDW 58wt%) and another with high wt% of PEGDA (6wt% PEGDA, 34wt% acrylic acid, TDW 58 wt%). For both printing compositions, 2wt% of the photoinitiator, Diphenyl(2,4,6-trimethylbenzoyl)phosphine oxide (TPO) in the form of water dispersible nanoparticles were added (with ionic SDS as a surfactant). All the materials were stirred together until the photoinitiator powder was fully dispersed and a clear solution was obtained. Rhodamine 6G was added as a dye to the solution with a high wt% of PEGDA, to distinguish between the solutions.

***Multi-material 3D printing:*** The inks were printed with a DLP printer (Pico2, Asiga, Australia) with a 385nm UV-LED light source. Each printed layer's thickness was 200 µm, with a light intensity of 25 mW/cm$^2$ and exposure times of 3 and 4 seconds, for the high and low PEGDA wt%, respectively. The dual-material printing was utilized by pausing the printing and switching the baths. In total, three bath replacements were conducted to form two multi-material layers.

## Authors contribution

I.L., E.Sh, and S.M. designed the research. I.L. designed the printed structures. E.Sa., R.L., and N.B. printed the hydrogel sheets. I.L. measured the induced shape transformations. I.L., E.Sa, E.Sh, and S.M. wrote the manuscript and produced the figures. E.Sh. and S.M. have jointly supervised the work.

## Data availability

The data that support the findings of this study are available from the corresponding author upon request.

## Competing interests

The authors declare no competing interests.

# References

1. Yoneda, M., Kobayakawa, Y., Kubota, H. Y. & Sakai, M. Surface contraction waves in amphibian eggs. *J Cell Sci* **54**, 35–46 (1982).

2. Nath, U., Crawford, B. C. W., Carpenter, R. & Coen, E. Genetic Control of Surface Curvature. *Science (1979)* **299**, 1404–1407 (2003).

3. Armon, S., Bull, M. S., Aranda-Diaz, A. & Prakash, M. Ultrafast epithelial contractions provide insights into contraction speed limits and tissue integrity. *Proceedings of the National Academy of Sciences* **115**, E10333–E10341 (2018).

4. Armon, S., Efrati, E., Kupferman, R. & Sharon, E. Geometry and Mechanics in the Opening of Chiral Seed Pods. *Science (1979)* **333**, 1726–1730 (2011).

5. Guest, S., Kebadze, E. & Pellegrino, S. A zero-stiffness elastic shell structure. *J Mech Mater Struct* **6**, 203–212 (2011).

6. Pezzulla, M., Smith, G. P., Nardinocchi, P. & Holmes, D. P. Geometry and mechanics of thin growing bilayers. *Soft Matter* **12**, 4435–4442 (2016).

7. Siéfert, E., Levin, I. & Sharon, E. Euclidean Frustrated Ribbons. *Phys Rev X* **11**, 011062 (2021).

8. Klein, Y., Efrati, E. & Sharon, E. Shaping of Elastic Sheets by Prescription of Non-Euclidean Metrics. *Science (1979)* **315**, 1116–1120 (2007).

9. McConney, M. E. *et al.* Topography from Topology: Photoinduced Surface Features Generated in Liquid Crystal Polymer Networks. *Advanced Materials* **25**, 5880–5885 (2013).

10. Aharoni, H., Xia, Y., Zhang, X., Kamien, R. D. & Yang, S. Universal inverse design of surfaces with thin nematic elastomer sheets. *Proceedings of the National Academy of Sciences* **115**, 7206–7211 (2018).

11. Kim, J., Hanna, J. A., Byun, M., Santangelo, C. D. & Hayward, R. C. Designing Responsive Buckled Surfaces by Halftone Gel Lithography. *Science (1979)* **335**, 1201–1205 (2012).

12. Wu, Z. L. *et al.* Three-dimensional shape transformations of hydrogel sheets induced by small-scale modulation of internal stresses. *Nat Commun* **4**, 1586 (2013).

13. Nojoomi, A., Jeon, J. & Yum, K. 2D material programming for 3D shaping. *Nat Commun* **12**, 603 (2021).

14. Hajiesmaili, E. & Clarke, D. R. Reconfigurable shape-morphing dielectric elastomers using spatially varying electric fields. *Nat Commun* **10**, 183 (2019).

15. Schild, H. G. Poly(N-isopropylacrylamide): experiment, theory and application. *Prog Polym Sci* **17**, 163–249 (1992).

16. Dong, Y. *et al.* 4D Printed Hydrogels: Fabrication, Materials, and Applications. *Adv Mater Technol* **5**, 2000034 (2020).

17. Hirokawa, Y. & Tanaka, T. Volume phase transition in a nonionic gel. *J Chem Phys* **81**, 6379–6380 (1984).

18. Banerjee, H., Suhail, M. & Ren, H. Hydrogel Actuators and Sensors for Biomedical Soft Robots: Brief Overview with Impending Challenges. *Biomimetics* **3**, 15 (2018).

19. Han, D. *et al.* Soft Robotic Manipulation and Locomotion with a 3D Printed Electroactive Hydrogel. *ACS Appl Mater Interfaces* **10**, 17512–17518 (2018).

20. Larush, L. *et al.* 3D printing of responsive hydrogels for drug-delivery systems. *J 3D Print Med* **1**, 219–229 (2017).

21. Pawar, A. A. *et al.* Rapid Three-Dimensional Printing in Water Using Semiconductor-Metal Hybrid Nanoparticles as Photoinitiators. *Nano Lett* **17**, 4497–4501 (2017).

22. Pawar, A. A. *et al.* High-performance 3D printing of hydrogels by water-dispersible photoinitiator nanoparticles. *Sci Adv* **2**, (2016).

23. Ge, Q. *et al.* 3D printing of highly stretchable hydrogel with diverse UV curable polymers. *Sci Adv* **7**, (2021).

24. Hua, M. *et al.* 4D Printable Tough and Thermoresponsive Hydrogels. *ACS Appl Mater Interfaces* **13**, 12689–12697 (2021).

25. Highley, C. B., Rodell, C. B. & Burdick, J. A. Direct 3D Printing of Shear-Thinning Hydrogels into Self-Healing Hydrogels. *Advanced Materials* **27**, 5075–5079 (2015).

26. Liu, S. & Li, L. Ultrastretchable and Self-Healing Double-Network Hydrogel for 3D Printing and Strain Sensor. *ACS Appl Mater Interfaces* **9**, 26429–26437 (2017).

27. Huang, L. *et al.* Ultrafast Digital Printing toward 4D Shape Changing Materials. *Advanced Materials* **29**, 1605390 (2017).

28. Sydney Gladman, A., Matsumoto, E. A., Nuzzo, R. G., Mahadevan, L. & Lewis, J. A. Biomimetic 4D printing. *Nat Mater* **15**, 413–418 (2016).

29. Zhou, Y., Duque, C. M., Santangelo, C. D. & Hayward, R. C. Biasing Buckling Direction in Shape-Programmable Hydrogel Sheets with Through-Thickness Gradients. *Adv Funct Mater* **29**, 1905273 (2019).

30. Boley, J. W. *et al.* Shape-shifting structured lattices via multimaterial 4D printing. *Proceedings of the National Academy of Sciences* **116**, 20856–20862 (2019).

31. Efrati, E., Sharon, E. & Kupferman, R. Elastic theory of unconstrained non-Euclidean plates. *J Mech Phys Solids* **57**, 762–775 (2009).

32. Klein, Y., Venkataramani, S. & Sharon, E. Experimental Study of Shape Transitions and Energy Scaling in Thin Non-Euclidean Plates. *Phys Rev Lett* **106**, 118303 (2011).

33. Huang, C., Wang, Z., Quinn, D., Suresh, S. & Hsia, K. J. Differential growth and shape formation in plant organs. *Proceedings of the National Academy of Sciences* **115**, 12359–12364 (2018).

34. Sharon, E., Roman, B. & Swinney, H. L. Geometrically driven wrinkling observed in free plastic sheets and leaves. *Phys Rev E* **75**, 046211 (2007).

35. Portet, T. *et al.* Ripples at edges of blooming lilies and torn plastic sheets. *Biophysj* **121**, 2389–2397 (2022).

36. Levin, I. & Sharon, E. Anomalously Soft Non-Euclidean Springs. *Phys Rev Lett* **116**, 035502 (2016).

37. Grossman, D., Sharon, E. & Diamant, H. Elasticity and Fluctuations of Frustrated Nanoribbons. *Phys Rev Lett* **116**, 258105 (2016).

38. Sun, K. & Mao, X. Fractional Excitations in Non-Euclidean Elastic Plates. *Phys Rev Lett* **127**, 98001 (2021).

39. Sawhney, R. & Crane, K. Boundary First Flattening. *ACM Trans Graph* **37**, 1–14 (2018).

40. Panetta, J. *et al.* Computational inverse design of surface-based inflatables. *ACM Trans Graph* **40**, 1–14 (2021).

41. Arieli, M., Moshe, M. & Sharon, E. Mechanical design principles in frustrated thin elastic sheets. *Soft Matter* **20**, 4414–4421 (2024).

42. Thérien-Aubin, H., Wu, Z. L., Nie, Z. & Kumacheva, E. Multiple shape transformations of composite hydrogel sheets. *J Am Chem Soc* **135**, 4834–4839 (2013).

43. Gu, J. *et al.* Geodesy: Self-rising 2.5D Tiles by Printing along 2D Geodesic Closed Path. *Conference on Human Factors in Computing Systems - Proceedings* 1–10 (2019) doi:10.1145/3290605.3300267.

44. Griniasty, I., Aharoni, H. & Efrati, E. Curved Geometries from Planar Director Fields: Solving the Two-Dimensional Inverse Problem. *Phys Rev Lett* **123**, 127801 (2019).

45. Timoshenko, S. Analysis of Bi-Metal Thermostats. *J Opt Soc Am* **11**, 233 (1925).

# Supplementary information for *4D Printing of Programmable Digital Metamaterials*

## Calculating the Gaussian curvature from a metric

As stated by *Gauss's Theorema Egregium*, the Gaussian curvature is invariant under isometries and, hence, can be calculated directly from the metric tensor.

For a general metric tensor, $\boldsymbol{a}(u,v) \equiv \begin{pmatrix} E(u,v) & F(u,v) \\ F(u,v) & G(u,v) \end{pmatrix}$, the calculation is cumbersome and can be expressed using the general Brioschi formula

$$K = \frac{1}{2}(\det \boldsymbol{a})^{-2}\left[\det\begin{pmatrix} 2F_{uv} - E_{vv} - G_{uu} & E_u & 2F_u - E_v \\ 2F_v - G_u & 2E & 2F \\ G_v & 2F & 2G \end{pmatrix} - \det\begin{pmatrix} 0 & E_v & G_u \\ E_v & 2E & 2F \\ G_u & 2F & 2G \end{pmatrix}\right]$$

, where subscript indices denote derivation with respect to the corresponding coordinates.

Since our gels swell uniformly from flat configurations, we can greatly simplify this expression by choosing to work with a coordinate system of the flat gel. Usually these are either the Cartesian or the polar coordinates depending on the symmetries of the desired swelling field.

For Cartesian coordinates ($x$ and $y$), $\bar{\boldsymbol{a}} \equiv \omega_0^2(x,y)\begin{pmatrix} 1 & 0 \\ 0 & 1 \end{pmatrix}$, and the programmed Gaussian curvature reads: $\bar{K} = -\frac{(\partial_x^2 + \partial_y^2)\omega_0^2}{2\omega_0^2}$.

In cases with polar symmetry, it is easier to work with the polar coordinates $r$ and $\varphi$, hence $\bar{\boldsymbol{a}} \equiv \omega_0^2(r)\begin{pmatrix} 1 & 0 \\ 0 & r^2 \end{pmatrix}$, and the programmed Gaussian curvature reads: $\bar{K} = \frac{r\omega_0'^2 - \omega_0(\omega_0' + r\omega_0'')}{r\omega_0^4}$.

The same expressions are used to calculate the needed differential swelling field, $\omega_0$ for a desired shape with Gaussian curvature, $\bar{K}$, by solving the differential equations above.

## The reduction of the growth rule to metric and curvature

Introducing a general swelling field, $\omega(z)$, to a voxel-sized, $\ell_0$, will result in the swelling of the midline, $\omega_0$, and curvature $\kappa$. For $-\frac{t}{2} \leq z \leq \frac{t}{2}$ we get that $\bar{\ell}(z) = \ell_0\, \omega(z)$ and that $\ell(z) = \omega_0 \ell_0 (1 + +\kappa\, z)$.

The energy, of that voxel, is

$$E(z) \propto \left(\bar{\ell}(z) - \ell(z)\right)^2 \propto \left(\omega(z) - \omega_0(1 + \kappa\, z)\right)^2 = \omega^2(z) - 2\omega(z)\omega_0(1 + \kappa\, z) + \omega_0^2(1 + \kappa\, z)^2$$

and we minimize $E \equiv \int E(z)dz$, therefore

$$0=\partial_{\omega_0}E\propto\int dz[-2\omega(z)(1+\kappa\, z)+2\omega_0(1+\kappa\, z)^2]$$

hence

$$\int dz\,\omega(z)(1+\kappa\, z)\;=\int dz\,\omega_0(1+\kappa\, z)^2\;=\omega_0(\,t+\frac{\kappa^2}{12}t^3)$$

Also,

$$0=\partial_\kappa E\propto\int dz[-2\omega(z)\omega_0\, z+2\omega_0^2(z+\kappa z^2)]$$

hence

$$\int dz\,\omega(z)\,z=\int dz\,\omega_0(z+\kappa z^2)=\omega_0\frac{\kappa}{12}t^3$$

then

$$\int dz\,\omega(z)\;=\omega_0\left(\,t+\frac{\kappa^2}{12}t^3\right)-\omega_0\frac{\kappa^2}{12}t^3=\omega_0 t$$

And finally

$$\omega_0=\frac{1}{t}\int dz\,\omega(z)=\langle\omega\rangle$$
$$\kappa=\frac{12}{\omega_0 t^3}\int dz\,\omega(z)\,z=\frac{12}{\omega_0 t^2}\langle\omega\, z\rangle$$

In our case $\omega(z)=\begin{cases}\omega_\mathrm{t} & z>0\\ \omega_\mathrm{b} & z\le 0\end{cases}$ and so $\begin{cases}\omega_0=\frac{\omega_\mathrm{t}+\omega_\mathrm{b}}{2}\\ \kappa=\frac{12}{\omega_0 t^3}\frac{t^2}{8}(\omega_\mathrm{t}-\omega_\mathrm{b})=\frac{3}{t}\frac{\omega_\mathrm{t}-\omega_\mathrm{b}}{\omega_\mathrm{t}+\omega_\mathrm{b}}\end{cases}$

The latter coincides with Timoshenko's result of $\kappa=\frac{3\epsilon}{2t}$ (for $\epsilon=\frac{\omega_\mathrm{t}-\omega_\mathrm{b}}{(\omega_\mathrm{t}+\omega_\mathrm{b})/2}$)[45].

## Calculating $\omega_0$ and $\Delta$ for a spherical cap

We are looking for a metric $\bar{\boldsymbol{a}}$ with an associated Gaussian curvature $\bar{K}\equiv R^{-2}$, where $R$ is the desired radius of the spherical cap. Utilizing the inherent symmetry of the problem, we can work in polar coordinates and assume $\omega_0$ has no azimuthal dependence, hence, $\bar{\boldsymbol{a}}\equiv\omega_0^2(r)\begin{pmatrix}1 & 0\\ 0 & r^2\end{pmatrix}$.

Using the Brioschi formula for orthogonal metric (see section above), $\bar{K}=\frac{r\omega_0'^2-\omega_0(\omega_0'+r\omega_0'')}{r\omega_0^4}$, solving this differential equation yields $\omega_0(r)=\frac{R\,c_1\mathrm{sech}\,(c_1(\log r+R\,c_2))}{r}$, where $c_1$ and $c_2$ are constants of integration.

To get a spherical cap and to void a sharp casp at $r \to 0$, we choose $c_1 = 1$, and continue to determine $c_2$ such as $\omega_0(r = 0) = 1.25$ (the maximal swelling possible).

Then, the maximal radius of the printed disc is given by $\omega_0(r_{\text{max}}) = 1$ (the minimal swelling possible).